\documentstyle[prd,aps,preprint,tighten,epsfig]{revtex}
\begin{document}

\draft

\title{On the neutrino mass ordering and flavor mixing structure}
\author{{\bf Zhi-zhong Xing} \thanks{E-mail:
xingzz@ihep.ac.cn}}
\address{
Institute of High Energy Physics, Chinese Academy of
Sciences, Beijing 100049, China \\
}

\maketitle

\begin{abstract}
Whether the neutrino mass spectrum is normal ($m^{}_1 < m^{}_2 <
m^{}_3$) or abnormal ($m^{}_3 < m^{}_1 < m^{}_2$) remains an open
question, but we show that the latter possibility looks quite
unnatural when it is related to the lepton flavor mixing matrix $U$
in a way similar to the reasonable correlation between the quark
mass spectrum and the quark flavor mixing matrix. Taking into
account the freedom in choosing the basis of weak interactions, we
make a novel prediction for $|U^{}_{\tau 1}|/|U^{}_{\tau 2}|$ in
terms of the three neutrino masses in the large tau mass limit. This
result is testable, and it implies that the normal neutrino mass
ordering is more likely to coincide with the observed structure of
$U$.
\end{abstract}


\newpage

In a straightforward extension of the standard electroweak model
which allows its three neutrinos to be massive, a nontrivial
mismatch between the mass and flavor eigenstates of leptons or
quarks arises from the fact that lepton or quark fields can interact
with both scalar and gauge fields, leading to the puzzling phenomena
of flavor mixing and CP violation \cite{XZ}. The $3\times 3$ lepton
and quark flavor mixing matrices appearing in the weak
charged-current interactions are referred to, respectively, as the
Pontecorvo-Maki-Nakagawa-Sakata (PMNS) matrix $U$ \cite{PMNS} and
the Cabibbo-Kobayashi-Maskawa (CKM) matrix $V$ \cite{CKM}:
\begin{eqnarray}
-{\cal L}^{}_{\rm cc} & = & \frac{g}{\sqrt{2}} \left[
\overline{\left(e ~~ \mu ~~ \tau \right)^{}_{\rm L}} \ \gamma^\mu \
U \left(\matrix{\nu^{}_1 \cr \nu^{}_2 \cr \nu^{}_3
\cr}\right)^{}_{\rm L} W^-_\mu ~ + ~ \overline{\left(u ~~ c ~~ t
\right)^{}_{\rm L}} \ \gamma^\mu \ V \left(\matrix{d \cr s \cr b
\cr}\right)^{}_{\rm L} W^+_\mu \right] + {\rm h.c.}
\end{eqnarray}
with all the fermion fields being the mass eigenstates. By
convention $U$ and $V$ are defined to be associated with $W^-$ and
$W^+$, respectively. In Eq. (1) the charged leptons and the quarks
with the same electric charges all have the normal mass hierarchies
(i.e., $m^{}_e \ll m^{}_\mu \ll m^{}_\tau$, $m^{}_u \ll m^{}_c \ll
m^{}_t$ and $m^{}_d \ll m^{}_s \ll m^{}_b$ \cite{PDG}). Yet it
remains unclear whether the three neutrinos also have a normal mass
ordering ($m^{}_1 < m^{}_2 < m^{}_3$). Now that $m^{}_1 < m^{}_2$
has been fixed from solar neutrino oscillations \cite{Hagiwara}, the
only possible ``abnormal" mass ordering is $m^{}_3 < m^{}_1 <
m^{}_2$. The neutrino mass ordering is one of the central concerns
in particle physics, and it will be determined in the foreseeable
future with the help of either an accelerator-based neutrino
oscillation experiment (e.g., T2K and NO$\nu$A \cite{NOVA}) or a
reactor-based antineutrino oscillation experiment (e.g., JUNO
\cite{JUNO}), or both of them. Depending on the neutrino mass
ordering to be normal or abnormal, a number of physical processes
such as the neutrinoless double-beta decay will have remarkably
different observable consequences. If $m^{}_3 < m^{}_1 < m^{}_2$
turned out to be true in nature, one would have to explain what is
behind it at a fundamental level.

On the other hand, the observed pattern of the PMNS matrix $U$ is
very different from that of the CKM matrix $V$. The latter exhibits
an approximate symmetry about its diagonal axis and an obvious
hierarchy among its nine elements (i.e., $|V^{}_{tb}| > |V^{}_{ud}|
> |V^{}_{cs}| \gg |V^{}_{us}| > |V^{}_{cd}| \gg |V^{}_{cb}| >
|V^{}_{ts}| \gg |V^{}_{td}| > |V^{}_{ub}|$ \cite{Xing12}), while the
former is structurally less symmetrical and hierarchical. This
difference might be attributed to a non-hierarchical neutrino mass
spectrum, or distinct underlying flavor symmetries which are
separately associated with leptons and quarks. In the lack of a full
flavor theory, it is a big challenge to resolve this kind of
problem. Some great ideas such as grand unifications,
supersymmetries and extra dimensions are not very helpful to deal
with most of the flavor puzzles, and current exercises of various
group languages or flavor symmetries are too divergent to converge
to something unique \cite{SV}. Although an access to the quark
flavor mixing structure seems possible in certain quark mass limits
\cite{X12}, no prediction in regard to it has been ventured.

In this work we show that the abnormal neutrino mass ordering
$m^{}_3 < m^{}_1 < m^{}_2$ looks quite unnatural when it is related
to the observed pattern of lepton flavor mixing in a way similar to
the reasonable correlation between the quark mass spectrum and the
CKM matrix. Taking account of the freedom in choosing the basis of
weak interactions, we predict
\begin{eqnarray}
\left|\frac{U^{}_{\tau 1}}{U^{}_{\tau 2}}\right| =
\sqrt{\frac{m^{}_1} {m^{}_2} \cdot \frac{m^{}_3 + m^{}_1}{m^{}_3 -
m^{}_1} \cdot \frac{m^{}_3 + m^{}_2}{m^{}_3 - m^{}_2}}
\end{eqnarray}
in the large tau mass limit. This novel relationship is more likely
to coincide with the present and future experimental data if
the neutrinos have a normal mass ordering $m^{}_1 < m^{}_2 < m^{}_3$.

Given the abnormal neutrino mass ordering $m^{}_3 < m^{}_1 <
m^{}_2$, one may ``renormalize" it to $m^\prime_1 < m^\prime_2 <
m^\prime_3$ by setting $m^\prime_1 = m^{}_3$, $m^\prime_2 = m^{}_1$
and $m^\prime_3 = m^{}_2$. In this case the corresponding neutrino
mass eigenstates are $\nu^\prime_1 = \nu^{}_3$, $\nu^\prime_2 =
\nu^{}_1$ and $\nu^\prime_3 = \nu^{}_2$. Then the weak
charged-current interactions in Eq. (1) can be rewritten as
\begin{eqnarray}
-{\cal L}^\prime_{\rm cc} = \frac{g}{\sqrt{2}} \left[
\overline{\left(e ~~ \mu ~~ \tau \right)^{}_{\rm L}} \ \gamma^\mu \
U^\prime \left(\matrix{\nu^\prime_1 \cr \nu^\prime_2 \cr
\nu^\prime_3 \cr}\right)^{}_{\rm L} W^-_\mu ~ + ~ \overline{\left(u
~~ c ~~ t \right)^{}_{\rm L}} \ \gamma^\mu \ V \left(\matrix{d \cr s
\cr b \cr}\right)^{}_{\rm L} W^+_\mu \right] + {\rm h.c.}
\end{eqnarray}
in the normal neutrino mass ordering (i.e., $m^\prime_1 < m^\prime_2
< m^\prime_3$) basis, where $U^\prime$ comes from $U$ through the
reordering transformation $(\nu^{}_1, \nu^{}_2, \nu^{}_3) \to
(\nu^\prime_2, \nu^\prime_3, \nu^\prime_1)$. As a result,
\begin{eqnarray}
U^\prime = \left( \matrix{
U^\prime_{e1} & U^\prime_{e2} & U^\prime_{e3} \cr
U^\prime_{\mu 1} & U^\prime_{\mu 2} & U^\prime_{\mu 3} \cr
U^\prime_{\tau 1} & U^\prime_{\tau 2} & U^\prime_{\tau 3} \cr}
\right) = U \left( \matrix{ 0 & 1 & 0 \cr 0 & 0 & 1 \cr 1 & 0 & 0 \cr}
\right) = \left( \matrix{
U^{}_{e3} & U^{}_{e1} & U^{}_{e2} \cr
U^{}_{\mu 3} & U^{}_{\mu 1} & U^{}_{\mu 2} \cr
U^{}_{\tau 3} & U^{}_{\tau 1} & U^{}_{\tau 2} \cr}
\right) \; .
\end{eqnarray}
A global $\chi^2$ analysis of those currently available neutrino
oscillation data \cite{FIT} allows us to obtain the magnitudes of
nine elements of $U^\prime$ at the $3\sigma$ confidence level:
\begin{eqnarray}
\left|U^\prime\right| = \left( \matrix{
0.126 \to 0.178 & ~0.795 \to 0.846~ & 0.513 \to 0.585 \cr
0.579 \to 0.808 & 0.205 \to 0.543 & 0.416 \to 0.730 \cr
0.567 \to 0.800 & 0.215 \to 0.548 & 0.409 \to 0.725 \cr} \right) \; ,
\end{eqnarray}
in which the unitarity of $U^\prime$ has been assumed and thus its
elements are correlated with one another. The smallest element of
$U^\prime$ is located on its upper left corner, while that of $U$ is
located on its upper right corner. This structural difference
implies that the flavor mixing angles of $U^\prime$ must be very
different from those of $U$ in a given parametrization. Let us adopt
the ``standard" parametrization advocated by the Particle Data Group
\cite{PDG},
\begin{eqnarray}
U = \left( \matrix{ c^{}_{12} c^{}_{13} & s^{}_{12} c^{}_{13} &
s^{}_{13} e^{-{\rm i} \delta} \cr -s^{}_{12} c^{}_{23} - c^{}_{12}
s^{}_{13} s^{}_{23} e^{{\rm i} \delta} & c^{}_{12} c^{}_{23} -
s^{}_{12} s^{}_{13} s^{}_{23} e^{{\rm i} \delta} & c^{}_{13}
s^{}_{23} \cr s^{}_{12} s^{}_{23} - c^{}_{12} s^{}_{13} c^{}_{23}
e^{{\rm i} \delta} & -c^{}_{12} s^{}_{23} - s^{}_{12} s^{}_{13}
c^{}_{23} e^{{\rm i} \delta} & c^{}_{13} c^{}_{23} \cr} \right)
P^{}_\nu \; ,
\end{eqnarray}
where $c^{}_{ij} \equiv \cos\theta^{}_{ij}$ and $s^{}_{ij} \equiv
\sin\theta^{}_{ij}$ (for $ij = 12, 13, 23$), and $P^{}_\nu = {\rm
Diag}\{e^{{\rm i} \rho}, e^{{\rm i} \sigma}, 1\}$ denotes the
Majorana phase matrix provided three massive neutrinos are the
Majorana particles. $U^\prime$ takes the same parametrization with
$P^\prime_\nu = {\rm Diag}\{1, e^{{\rm i} \rho}, e^{{\rm i}
\sigma}\}$ coming from $P^{}_\nu$ through the reordering
transformation $(\nu^{}_1, \nu^{}_2, \nu^{}_3) \to (\nu^\prime_2,
\nu^\prime_3, \nu^\prime_1)$. The exact analytical relations between
the two sets of flavor mixing parameters of $U$ and $U^\prime$ are
given by
\begin{eqnarray}
t^\prime_{12} &= & \left| \frac{U^\prime_{e2}}{U^\prime_{e1}}
\right| = \left| \frac{U^{}_{e1}}{U^{}_{e3}} \right| =
\frac{c^{}_{12} c^{}_{13}}{s^{}_{13}} \; ,
\nonumber \\
s^\prime_{13} & = & \left| U^\prime_{e3} \right| = \left| U^{}_{e2}
\right| = s^{}_{12} c^{}_{13} \; ,
\nonumber \\
t^\prime_{23} &= & \left| \frac{U^\prime_{\mu 3}}{U^\prime_{\tau 3}}
\right| = \left| \frac{U^{}_{\mu 2}}{U^{}_{\tau 2}} \right| =
\frac{\sqrt{c^2_{12} c^2_{23} - 2 c^{}_{12} s^{}_{12} s^{}_{13}
c^{}_{23} s^{}_{23} c^{}_{\delta} + s^2_{12} s^2_{13} s^2_{23}}}
{\sqrt{c^2_{12} s^2_{23} + 2 c^{}_{12} s^{}_{12} s^{}_{13} c^{}_{23}
s^{}_{23} c^{}_{\delta} + s^2_{12} s^2_{13} c^2_{23}}} \; ,
\nonumber \\
s^\prime_{\delta} & = & \frac{{\rm Im}\left( U^\prime_{e2}
U^\prime_{\mu 3} U^{\prime *}_{e 3} U^{\prime *}_{\mu 2}\right)}
{c^\prime_{12} s^\prime_{12} c^{\prime 2}_{13} s^\prime_{13}
c^\prime_{23} s^\prime_{23}} = \frac{{\rm Im}\left( U^{}_{e1}
U^{}_{\mu 2} U^*_{e 2} U^*_{\mu 1}\right)} {c^\prime_{12}
s^\prime_{12} c^{\prime 2}_{13} s^\prime_{13} c^\prime_{23}
s^\prime_{23}} = \frac{ c^{}_{12} s^{}_{12} c^2_{13} s^{}_{13}
c^{}_{23} s^{}_{23}} {c^\prime_{12} s^\prime_{12} c^{\prime 2}_{13}
s^\prime_{13} c^\prime_{23} s^\prime_{23}} s^{}_{\delta} \; ,
\end{eqnarray}
where $t^\prime_{ij} \equiv \tan\theta^\prime_{ij}$, $c^{}_{\delta}
\equiv \cos\delta$, $s^{}_{\delta} \equiv \sin\delta$ and
$s^\prime_{\delta} \equiv \sin\delta^\prime$. In view of Eq. (5)
or Ref. \cite{FIT} with $\delta \in [0^\circ, 360^\circ]$ and
$\delta^\prime \in [0^\circ, 360^\circ]$, we arrive at the $3\sigma$
ranges of three flavor mixing angles as
\begin{eqnarray}
\theta^\prime_{12} & = & 77.4^\circ \to 81.5^\circ \; , ~~~
\theta^\prime_{13} = 30.9^\circ \to 35.8^\circ \; , ~~~
\theta^\prime_{23} = 29.8^\circ \to 60.7^\circ \; ;
\nonumber \\
\theta^{}_{12} & = & 31.1^\circ \to 35.9^\circ \; , ~~~
\theta^{}_{13} = 7.2^\circ \to 10.0^\circ \; , ~~\;\;\;
\theta^{}_{23} = 35.8^\circ \to 54.8^\circ \; .
\end{eqnarray}
In comparison, the three quark flavor mixing angles in the same
parametrization of the CKM matrix $V$ read $\vartheta^{}_{12} =
13.023^\circ \pm 0.038^\circ$, $\vartheta^{}_{13} =
0.201^{+0.009^\circ}_{-0.008^\circ}$ and $\vartheta^{}_{23} =
2.361^{+0.063^\circ}_{-0.028^\circ}$, extracted from current
experimental data as given by \cite{PDG}
\begin{eqnarray}
|V| = \left(\matrix{ 0.97427 \pm 0.00015 & 0.22534 \pm 0.00065 &
0.00351^{+0.00015}_{-0.00014} \cr 0.22520 \pm 0.00065 & 0.97344 \pm
0.00016 & 0.0412^{+0.0011}_{-0.0005} \cr
0.00867^{+0.00029}_{-0.00031} & 0.0404^{+0.0011}_{-0.0005} &
0.999146^{+0.000021}_{-0.000046} \cr} \right) \; .
\end{eqnarray}
The CP-violating phase of $V$, denoted as $\delta^{}_q$ in this
parametrization, has also been determined to a good degree of
accuracy: $\delta^{}_q = 69.21^{+2.55^\circ}_{-4.59^\circ}$
\cite{PDG}. It seems very hard to find any potentially interesting
numerical correlation between the three lepton mixing parameters
$(\theta^\prime_{12}, \theta^\prime_{13}, \theta^\prime_{23})$ and
the three quark mixing parameters $(\vartheta^{}_{12},
\vartheta^{}_{13}, \vartheta^{}_{23})$, and the previous
phenomenological conjectures $\theta^{}_{12} + \vartheta^{}_{12} =
45^\circ$ and $\theta^{}_{23} \pm \vartheta^{}_{23} = 45^\circ$ are
subject to the parametrization (or the flavor basis) itself and
maybe instable against the renormalization-group running effects
\cite{Xing05}. Although one may similarly make the conjectures like
$\theta^\prime_{12} + \vartheta^{}_{12} = 90^\circ$ and
$\theta^\prime_{23} \pm \vartheta^{}_{23} = 45^\circ$, they are
unlikely to provide us with any enlightening information about the
underlying dynamics of lepton and quark flavor mixing.

Compared with the normal mass hierarchies of up- and down-type
quarks which are associated with a hierarchical structure of the CKM
matrix $V$, the ``normalized" neutrino mass ordering $m^\prime_1 <
m^\prime_2 < m^\prime_3$ corresponds to a quite unnatural structure
of the PMNS matrix $U^\prime$ as illustrated in Eq. (5). The normal
neutrino mass ordering $m^{}_1 < m^{}_2 < m^{}_3$ is favored in this
connection, because its corresponding PMNS matrix $U$ has a somewhat
more natural structure \cite{FIT}
\begin{eqnarray}
\left|U\right| = \left( \matrix{ 0.795 \to 0.846 & ~0.513 \to 0.585~
& 0.126 \to 0.178 \cr 0.205 \to 0.543 & 0.416 \to 0.730 & 0.579 \to
0.808 \cr 0.215 \to 0.548 & 0.409 \to 0.725 & 0.567 \to 0.800 \cr}
\right) \; .
\end{eqnarray}
This kind of ``naturalness" might come from a comparison between $U$
and $V$, in contrast to a comparison between $U^\prime$ and $V$.
However, one should keep in mind that the definitions of $U$ and $V$
are related to $W^-$ and $W^+$, respectively. If ${\cal L}^{}_{\rm
cc}$ in Eq. (1) is rewritten as
\begin{eqnarray}
-{\cal L}^{}_{\rm cc} = \frac{g}{\sqrt{2}} ~\overline{\left(e ~~ \mu
~~ \tau ~~ d ~~ s ~~ b \right)^{}_{\rm L}} \ \gamma^\mu \left(
\matrix{U & {\bf 0} \cr {\bf 0} & V^\dagger \cr} \right)
\left(\matrix{\nu^{}_1 \cr \nu^{}_2 \cr \nu^{}_3 \cr u \cr c \cr t
\cr} \right)^{}_{\rm L} W^-_\mu + {\rm h.c.} \; ,
\end{eqnarray}
then it seems more reasonable to compare between $U$ and $V^\dagger$
instead of $V$ itself. But the magnitudes of six off-diagonal
elements of $V$ exhibit an approximate symmetry about its diagonal
axis, and therefore $|V^\dagger| \simeq |V|$ is actually a good
approximation. Defining $V^\prime \equiv V^\dagger$ and taking the
same ``standard" parametrization for it, we obtain
$\vartheta^{\prime}_{12} = 13.015^\circ \pm 0.038^\circ$,
$\vartheta^{\prime}_{13} = 0.497^{+0.016^\circ}_{-0.018^\circ}$,
$\vartheta^{\prime}_{23} = 2.315^{+0.064^\circ}_{-0.028^\circ}$ and
$\delta^\prime_q = -22.23^{+1.33^\circ}_{-1.11^\circ}$. It is again
difficult to see any meaningful correlation between $U$ and
$V^\prime$ or between $U^\prime$ and $V^\prime$.

Now let us turn to the mass matrices of leptons and quarks so as to
look at their possible relations with the observed structures of $U$
and $V$. Assuming massive neutrinos to be the Majorana particles, we
write the lepton and quark mass terms as
\begin{eqnarray}
-{\cal L}^{}_{\rm mass} = \overline{{\bf E}^{}_{\rm L}} M^{}_\ell
{\bf E}^{}_{\rm R} + \frac{1}{2} \overline{{\bf N}^{}_{\rm L}}
M^{}_\nu {\bf N}^{c}_{\rm L} + \overline{{\bf U}^{}_{\rm L}}
M^{}_{\rm u} {\bf U}^{}_{\rm R} + \overline{{\bf D}^{}_{\rm L}}
M^{}_{\rm d} {\bf D}^{}_{\rm R} + {\rm h.c.} \; ,
\end{eqnarray}
where ${\bf E}$, ${\bf U}$ and ${\bf D}$ are the column vectors of
charged leptons, up- and down-type quarks, respectively; ${\bf
N}^{c}_{\rm L}$ represents the charge conjugation of the column
vector of three left-handed neutrinos ${\bf N}^{}_{\rm L}$; and the
effective Majorana neutrino mass matrix $M^{}_\nu$ is symmetrical.
Because the standard weak interactions do not involve any
flavor-changing right-handed currents, the relevant physics keeps
unchanged if each right-handed column vector in Eq. (12) undergoes
an arbitrary unitary transformation. Without loss of generality,
this kind of freedom allows us to do two things: (a) making
$M^{}_\ell$, $M^{}_{\rm u}$ and $M^{}_{\rm d}$ all Hermitian
\cite{Frampton}; (b) obtaining three zeros for $M^{}_{\rm u}$ and
$M^{}_{\rm d}$ in a suitable flavor basis, and similarly three zeros
for $M^{}_\ell$ and $M^{}_\nu$ (by convention, a pair of
off-diagonal zeros in a Hermitian or symmetrical fermion mass matrix
is always counted as one zero) \cite{Branco}. Here we work in the
basis where both $M^{}_\nu$ and $M^{}_{\rm d}$ take the Fritzsch
texture \cite{F78}, while $M^{}_\ell$ and $M^{}_{\rm u}$ are
arbitrary Hermitian matrices:
\begin{eqnarray}
M^{}_{\rm u} = \left( \matrix{E^{}_{\rm u} & C^{}_{\rm u} &
F^{}_{\rm u} \cr C^*_{\rm u}  & D^{}_{\rm u}  & B^{}_{\rm u} \cr
F^*_{\rm u}  & B^*_{\rm u}  & A^{}_{\rm u} \cr} \right) \; ,  &~~~~&
M^{}_{\rm d} = \left( \matrix{ 0 & C^{}_{\rm d} & 0 \cr C^{*}_{\rm
d} & 0 & B^{}_{\rm d} \cr 0 & B^{*}_{\rm d} & A^{}_{\rm d} \cr}
\right) \; ;
\nonumber \\
M^{}_\ell = \left( \matrix{E^{}_\ell & C^{}_\ell & F^{}_\ell \cr
C^*_\ell & D^{}_\ell & B^{}_\ell \cr F^*_\ell & B^*_\ell & A^{}_\ell \cr}
\right) \; , &~~~~& M^{}_\nu = \left( \matrix{ 0 & C^{}_\nu & 0 \cr
C^{}_\nu & 0 & B^{}_\nu \cr 0 & B^{}_\nu & A^{}_\nu \cr}
\right) \; .
\end{eqnarray}
It is worth stressing that the three zeros in either $M^{}_{\rm d}$
or $M^{}_\nu$ come from a proper choice of the flavor basis and do
not have any physical content by themselves, but any more texture
zeros in the lepton or quark sector must be subject to a
phenomenological assumption or some kind of model dependence
\cite{FX00}. Defining $H^{}_x \equiv M^{}_x M^\dagger_x$ (for $x =
{\rm u}, {\rm d}$ or $\ell$), one may diagonalize it by means of a
unitary transformation $O^\dagger_x H^{}_x O^{}_x =
\widehat{H}^{}_x$ (i.e., $\widehat{H}^{}_{\rm u} = {\rm
Diag}\{m^2_u, m^2_c, m^2_t\}$, $\widehat{H}^{}_{\rm d} = {\rm
Diag}\{m^2_d, m^2_s, m^2_b\}$ or $\widehat{H}^{}_\ell = {\rm
Diag}\{m^2_e, m^2_\mu, m^2_\tau\}$). Furthermore, the Majorana
neutrino mass matrix $M^{}_\nu$ can be diagonalized via the
transformation $O^\dagger_\nu M^{}_\nu O^*_\nu = \widehat{M}^{}_\nu
= {\rm Diag}\{ m^{}_1, m^{}_2, m^{}_3\}$. Thanks to the Fritzsch
form of $M^{}_{\rm d}$ or $M^{}_\nu$, the elements of $O^{}_{\rm d}$
or $O^{}_\nu$ can be exactly calculated in terms of the mass ratios
of three down-type quarks or three neutrinos \cite{Georgi}. The main
issue for now is how to deal with $M^{}_\ell$ or $M^{}_{\rm u}$,
such that the elements of the PMNS matrix $U = O^\dagger_\ell
O^{}_\nu$ and those of the CKM matrix $V = O^\dagger_{\rm u}
O^{}_{\rm d}$ can be fully or partially calculated via
\begin{eqnarray}
U^{}_{\alpha i} = \sum^3_{k=1}
(O^{}_\ell)^*_{k \alpha} (O^{}_\nu)^{}_{k i} \; , ~~~~~~~~
V^{}_{\alpha i} = \sum^3_{k=1}
(O^{}_{\rm u})^*_{k \alpha} (O^{}_{\rm d})^{}_{k i} \; ,
\end{eqnarray}
in which the subscripts $\alpha$ and $i$ run over $(e, \mu, \tau)$
and $(1, 2, 3)$ for $U$ or over $(u, c, t)$ and $(d, s, b)$ for $V$,
respectively. The possibilities of assuming extra zeros in or
imposing a certain flavor symmetry on the textures of $M^{}_\ell$
and $M^{}_{\rm u}$ have already been discussed in the literature
\cite{FX00}. But is there any other way to proceed?

We find that it should be a reasonable approximation to deal with
$H^{}_{\rm u}$ (or $H^{}_\ell$) in the large top (or tau) mass
limit, simply because $m^{}_t$ (or $m^{}_\tau$) is the largest mass
among the six quark (or lepton) masses. In the $m^{}_t \to \infty$
limit the top quark is expected to be essentially decoupled from the
up and charm quarks in $H^{}_{\rm u}$ \cite{F87}, and a similar
situation exists in $H^{}_\ell$ in the $m^{}_\tau \to \infty$ limit.
Therefore, we have
\begin{eqnarray}
\lim_{m^{}_t \to \infty} H^{}_{\rm u} \propto \left(
\matrix{\times & \times & 0 \cr \times & \times & 0 \cr 0 & 0 &
\infty \cr} \right) \; , &~~~~&
\lim_{m^{}_t \to \infty} O^{}_{\rm u} \propto \left(
\matrix{\times & \times ~& 0 \cr \times & \times & 0 \cr 0 & 0 & 1
\cr} \right) \; ,
\nonumber \\
\lim_{m^{}_\tau \to \infty} H^{}_\ell \propto \left(
\matrix{\times & \times & 0 \cr \times & \times & 0 \cr 0 & 0 &
\infty \cr} \right) \; , &~~~~&
\lim_{m^{}_\tau \to \infty} O^{}_\ell \propto \left(
\matrix{\times & \times ~& 0 \cr \times & \times & 0 \cr 0 & 0 & 1
\cr} \right) \; ,
\end{eqnarray}
where the symbol ``$\times$" denotes a nonzero matrix element.
A combination of Eqs. (14) and (15) leads us to the novel predictions
\begin{eqnarray}
\lim_{m^{}_t \to \infty} \left|\frac{V^{}_{td}}{V^{}_{ts}}\right| =
\left|\frac{(O^{}_{\rm d})^{}_{3 d}}{(O^{}_{\rm d})^{}_{3 s}}\right|
= \sqrt{\frac{m^{}_d} {m^{}_s} \cdot \frac{m^{}_b + m^{}_d}{m^{}_b -
m^{}_d} \cdot \frac{m^{}_b + m^{}_s}{m^{}_b - m^{}_s}} \;\;\;
\end{eqnarray}
for the lower left corner of the CKM matrix $V$ and
\begin{eqnarray}
\lim_{m^{}_\tau \to \infty} \left|\frac{U^{}_{\tau 1}}{U^{}_{\tau
2}}\right| = \left|\frac{(O^{}_\nu)^{}_{3 1}}{(O^{}_\nu)^{}_{3
2}}\right| = \sqrt{\frac{m^{}_1} {m^{}_2} \cdot \frac{m^{}_3 +
m^{}_1}{m^{}_3 - m^{}_1} \cdot \frac{m^{}_3 + m^{}_2}{m^{}_3 -
m^{}_2}} \;\;\;
\end{eqnarray}
for the lower left corner of the PMNS matrix $U$. In obtaining Eqs.
(16) and (17), we have used the exact analytical expressions of
$O^{}_{\rm d}$ and $O^{}_\nu$ given in Ref. \cite{Georgi} for the
Fritzsch texture of $M^{}_{\rm d}$ and $M^{}_\nu$. Taking account of
the central values of $m^{}_d = 2.82 \pm 0.48$ MeV, $m^{}_s =
57^{+18}_{-12}$ MeV and $m^{}_b = 2.86^{+0.16}_{-0.06}$ GeV at the
electroweak scale \cite{XZZ}, we arrive at $|V^{}_{td}|/|V^{}_{ts}|
= 0.227$. Although its error bar is rather appreciable due to
uncertainties of the quark masses (in particular, the big error
associated with $m^{}_s$), this result is in good agreement with the
accurate experimental value $|V^{}_{td}|/|V^{}_{ts}| =
0.215^{+0.010}_{-0.013}$ \cite{PDG}. The phenomenological success of
Eq. (16) encourages us to conjecture that Eq. (17) is very likely to
make sense in revealing an underlying correlation between the
neutrino mass spectrum and the lepton flavor mixing structure. We
illustrate the dependence of $|U^{}_{\tau 1}|/|U^{}_{\tau 2}|$ on
the smallest neutrino mass $m^{}_1$ (normal mass ordering or NMO) or
$m^{}_3$ (abnormal mass ordering or AMO) in FIG. 1, where we have
input $\Delta m^2_{21} \equiv m^2_2 - m^2_1 = (7.00 \to 8.09) \times
10^{-5} ~{\rm eV}^2$ and $\Delta m^2_{31} \equiv m^2_3 - m^2_1 =
(2.276 \to 2.695) \times 10^{-3} ~{\rm eV}^2$ (NMO) or $\Delta
m^2_{32} \equiv m^2_3 - m^2_2 = (-2.649 \to -2.242) \times 10^{-3}
~{\rm eV}^2$ (AMO) at the $3\sigma$ level \cite{FIT}. Because of the
unknown CP-violating phase $\delta$ and some uncertainties
associated with the flavor mixing angles, the present experimental
bound on the left-hand side of Eq. (17) remains quite loose:
$|U^{}_{\tau 1}|/|U^{}_{\tau 2}| = 0.30 \to 1.34$ as indicated by
Eq. (10). In this case either the NMO (i.e., $m^{}_1 < m^{}_2 <
m^{}_3$) or the AMO (i.e., $m^{}_3 < m^{}_1 < m^{}_2$) can coincide
with current neutrino oscillation data, but the former is apparently
favored by the observed pattern of lepton flavor mixing, as shown in
FIG. 1. In the foreseeable future more precise experimental data
will test the validity of Eq. (17) and single out the correct
neutrino mass ordering. Some comments and discussions are in order.
\begin{itemize}
\item     In the AMO case $m^{}_1$ and $m^{}_2$ are nearly equal, and
thus $|U^{}_{\tau 1}|/|U^{}_{\tau 2}| \gtrsim 1$ holds, as one can clearly
see from FIG. 1. If the experimental result of
$|U^{}_{\tau 1}|/|U^{}_{\tau 2}|$ turns out to be less than one, then
Eq. (17) will definitely lead us to the NMO. FIG. 1 also tells us that
the possibility of $m^{}_1 \to 0$ has been ruled out by Eq. (17), and
the possibility of $m^{}_3 \to 0$ is still allowed but seems not to be
favored.

\item     In the standard parametrization of $U$, Eq. (17) establishes
a relationship between the unknown neutrino mass scale and the unknown
CP-violating phase $\delta$:
\begin{eqnarray}
\frac{m^{}_1} {m^{}_2} \cdot \frac{m^{}_3 + m^{}_1}{m^{}_3 - m^{}_1}
\cdot \frac{m^{}_3 + m^{}_2}{m^{}_3 - m^{}_2} =
\frac{s^2_{12} s^2_{23} - 2 c^{}_{12} s^2_{12} s^{}_{13}
c^{}_{23} s^{}_{23} c^{}_\delta + c^2_{12} s^2_{13} c^2_{23}}
{c^2_{12} s^2_{23} + 2 c^{}_{12} s^2_{12} s^{}_{13} c^{}_{23} s^{}_{23}
c^{}_\delta + s^2_{12} s^2_{13} c^2_{23}} \; .
\end{eqnarray}
Although this kind of correlation is subject to the large tau mass
limit and more likely a leading-order approximation, it deserves an
experimental test. A similar correlation in the quark sector
described by Eq. (16) {\it does} survive the current experimental
test.
\end{itemize}
Finally, let us emphasize that any kind of model building on the
flavor structures of leptons and quarks depends on the choice of a
specific flavor basis. The basis chosen in Eq. (13) is just such a
case, but it has nothing to do with the phenomenological
assumptions. It is the large top or tau mass limit taken in Eq. (15)
that makes the novel predictions in Eqs. (16) and (17) possible. If
the latter also proves to be a success in phenomenology, then one
will be well motivated to search for the underlying flavor dynamics.

In summary, we have tried to explore some possible implications of a
normal or abnormal neutrino mass ordering on the lepton flavor
mixing structure. We show that $m^{}_3 < m^{}_1 < m^{}_2$ looks
quite unnatural after it is ``renormalized" and related to the PMNS
matrix in a way similar to the reasonable correlation between the
quark mass spectrum and the CKM matrix. With the help of the freedom
in choosing the basis of weak interactions, we have made the
nontrivial predictions for $|V^{}_{td}|/|V^{}_{ts}|$ in terms of the
three down-type quark masses in the large top mass limit and
$|U^{}_{\tau 1}|/|U^{}_{\tau 2}|$ in terms of the three neutrino
masses in the large tau mass limit. We find that the former is in
good agreement with current experimental data, and the latter
provides us with a preliminary hint that the normal neutrino mass
ordering $m^{}_1 < m^{}_2 < m^{}_3$ is more likely to coincide with
the observed structure of $U$.

It is certainly a theoretical challenge to pin down what is behind
the normal or abnormal neutrino mass ordering \cite{Goodman} and
whether there is a definite correlation between the fermion mass
spectra and flavor mixing patterns. Our present attempt in this
connection remains quite limited, but it has led us to some
encouraging and interesting results. The underlying flavor theory,
which might be related to a certain flavor symmetry and its
spontaneous or explicit breaking mechanism, should finally give us a
dynamical reason for the phenomena of lepton and quark flavor mixing
and CP violation. But we are now following the opposite way to look
for such a fundamental theory from the bottom up.

\vspace{0.5cm}

I would like to thank C. Giunti, M. Goodman, Y.F. Li, S. Zhou and
Y.L. Zhou for useful discussions. I am especially indebted to Y.L.
Zhou for his kind help in plotting the figure in this paper. I am
also grateful to A. Studenikin for his invitation and hospitality
during the 16th Lomonosov Conference on Elementary Particle Physics
in Moscow, where part of this paper was written. My research was
supported in part by the National Natural Science Foundation of
China under grant No. 11135009.

\begin{figure*}
\vspace{1cm} \centering
\includegraphics[width=16cm]{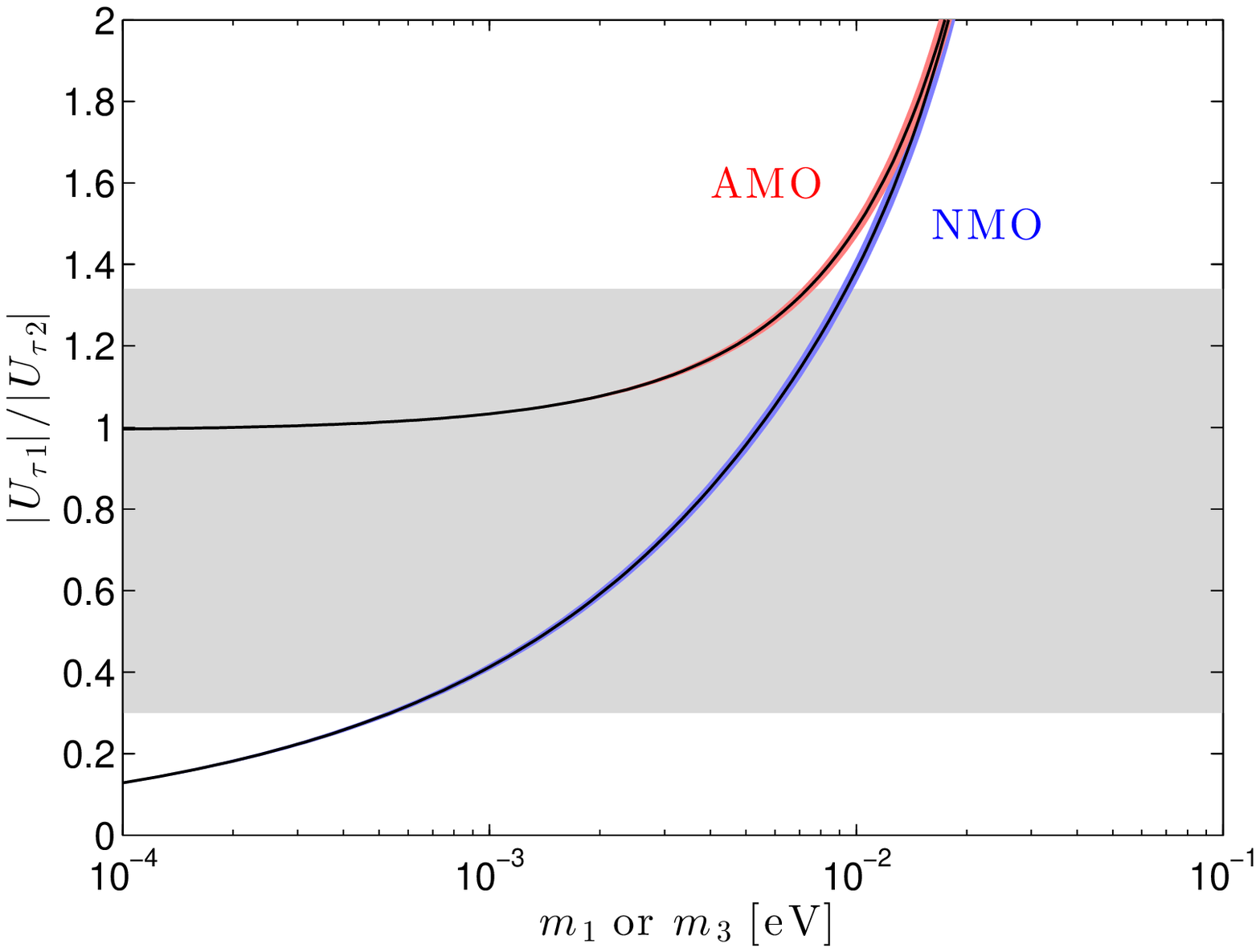}
\vspace{0.6cm} \caption{The dependence of $|U^{}_{\tau 1}|/U^{}_{\tau 2}|$
on the smallest neutrino mass $m^{}_1$ (normal mass ordering) or
$m^{}_3$ (abnormal mass ordering), as predicted by Eq. (17).
Here the red and blue regions of the two curves arise from the $3\sigma$
ranges of two neutrino mass-squared differences, and the grey band
stands for the range of $|U^{}_{\tau 1}|/U^{}_{\tau 2}|$ allowed by
current neutrino oscillation data [11].}
\end{figure*}

\end{document}